\documentclass[a4paper,10pt]{article}
\usepackage{latexsym}
\usepackage{amsmath}
\usepackage{amsfonts}
\usepackage{graphicx,amssymb}

\newcommand{\bbox}{\rule{2mm}{2mm}}
\newcommand{\cO}{\mathcal{O}}
\newcommand{\lan}{{\langle}}
\newcommand{\ran}{{\rangle}}
\newcommand{\cI}{{\cal I}}
\newcommand{\cE}{{\cal E}}
\newcommand{\cK}{{\cal K}}
\newcommand{\cT}{\mathit{{\cal {TO}}}}
\newcommand{\cW}{{\cal W}}
\newcommand{\cR}{{\cal R}}
\newcommand{\cP}{{\cal P}}
\newcommand{\cA}{{\cal A}}
\newcommand{\cS}{{\cal S}}
\newcommand{\cC}{{\cal C}}
\newcommand{\N}{{\mathbb N}}
\newcommand{\cM}{{\mathbb M}}
\newcommand{\bx}{{\Box}}
\newcommand{\dia}{{\Diamond}}
\newcommand{\bigc}{\bigcirc}

\newcommand{\true}{{\tt true}}
\newcommand{\false}{{\tt false}}
\newcommand{\next}{{\bigcirc}}
\newcommand{\unt}{{\cal U}}

\newcommand {\exs}{\exists}
\newcommand {\for}{\forall}

\newcommand{\cLT}{{\cal L}_{2}^{T}}
\newcommand{\cLTo}{{\cal L}_{2}^{TO}}
\newcommand{\LT}{{\cal L}^{T}}

\def \R {\mathbb{R}}
\def \nR {{\R}^{\geq 0}}
\def \Q {\mathbb{Q}}
\def \nQ {{\Q}^{\geq 0}}

\newtheorem{Theorem}{Theorem}
\newtheorem{Lemma}[Theorem]{Lemma}

\newtheorem{Fact}{Fact}

\newtheorem{Rule-def}[Theorem]{Rule}

\title{A Decidable Timeout based Extension of Propositional Linear Temporal Logic}
\author{Janardan Misra\\
     EMCSS India Pvt. Ltd., Bangalore 560048, India.\\ Email: janmishra@gmail.com \\
     \and Suman Roy\\
     SETLABS, Infosys Tech. Ltd., \#44 Electronic City, \\ Bangalore 560100, India.\\ Email: suman\_roy@infosys.com}

\begin{document}

\maketitle

\begin{abstract}
We develop a timeout based extension of propositional linear temporal logic (which we call TLTL) to specify timing properties of timeout based models of real time systems. TLTL formulas explicitly refer to a running global clock together with static timing variables as well as a dynamic variable abstracting the timeout behavior. 
We extend LTL with the capability to express timeout constraints. From the expressiveness view point, TLTL is not comparable with important known clock based real-time logics including TPTL, XCTL, and MTL, i.e., TLTL can specify certain properties, which cannot be specified in these logics (also vice-versa). We define a corresponding timeout tableau for satisfiability checking of the TLTL formulas. Also a model checking algorithm over timeout Kripke structure is presented. Further we prove that the validity checking for such an extended logic remains PSPACE-complete even in the presence of timeout constraints and infinite state models. Under discrete time semantics, with bounded timeout increments, the model-checking problem that if a TLTL-formula holds in a timeout Kripke structure is also PSPACE complete. We further prove that when TLTL is interpreted over discrete time, it can be embedded in the monadic second order logic with time, and 
when TLTL is interpreted over dense time without the condition of non-zenoness, the resulting logic becomes $\Sigma_1^1$-complete.
\end{abstract}

{\sf Keywords:} Timeout systems, Real time logics, Model checking, Timing properties, Timeout constraints, Tableau satisfiability, Undecidability

\section{Introduction}\label{intro}

Real-time systems are an important class of mission critical systems, which have been well studied for their design, implementation, and performance~\cite{OD08}. Designing faithful models for real-time systems essentially requires representing different kinds of timing behavior e.g., relative delays and timing constraints. In a timeout based design framework for real-time systems, 
timing requirements are modeled by defining the execution of an action in terms of an expiration of a delay, often represented as a timeout (or timer). Traditionally, timeouts have been used in real-time system designs for handling various timing scenarios including (forced) expiration of a waiting state. Dutertre and Sorea~\cite{DS04} used timeout based modeling to formally verify safety properties of the real-time systems with discrete dynamics. A timeout model contains a finite set of timeouts and a variable $x$ which keeps track of the current (global) time. Timeouts define the time points when discrete transitions are enabled in the future. In practice, a typical real-time system may contain $n$ concurrently active processes. Each process is associated with one timeout which denotes the future point of time when the next discrete transition for the corresponding process will occur. Transitions in this model are classified into two types - time progress transitions and discrete transitions. In a time progress transition, the time variable $x$ is advanced to the minimum valued timeout(s). A discrete transition occurs when $x$ is equal to the minimum valued timeout(s). If there are more than one processes, which have their timeouts equal to the minimum value, then some of them are randomly selected and corresponding discrete transitions take place with the values of the corresponding timeouts are set in the future.

\begin{figure}[ht]
\centering
\includegraphics[scale=0.6]{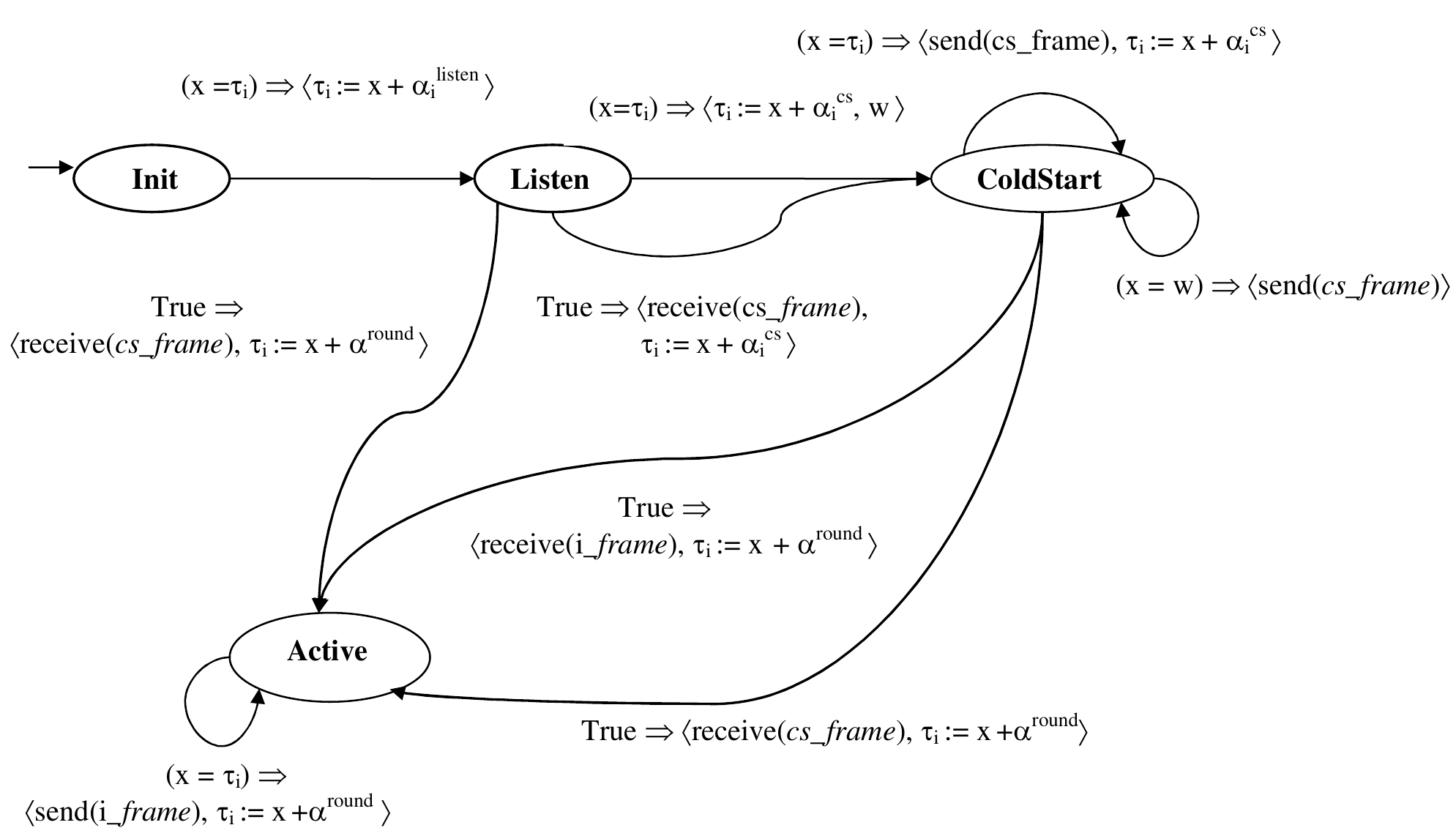}
\vspace{-0.4cm} \caption{State transition diagram of TTA startup algorithm at $i^{th}$ node. Edges are labeled as: 
{\sf guard}$\Rightarrow \lan${\sf [send/receive], timeout update, [record\_time\_var]}$\ran$, where (optional) {\sf record\_time\_var} records the time when a transition occurs on the edge.} \label{fig:fig1} 
\end{figure}

Startup algorithm for Time-triggered Architecture (TTA) is an example of a system where timeouts are explicitly used in the design. TTA start-up algorithm executes on a logical bus meant for safety critical application in both automotive and aerospace industries. In a normal operation, $N$ nodes share a TTA bus using a TDMA schedule. The state-machine of the startup algorithm executed on the nodes is shown in Figure~\ref{fig:fig1}. Each node $i \in [1,  N]$ has a local timeout $\tau_i$. Timeout increments in various states are defined in terms of timeout increment parameters: $\alpha_i^{listen} = (2N + i -1)\lambda$, $\alpha_i^{cs} = (N + i -1)\lambda$, and $\alpha_i^{round} = N\lambda$, where $\lambda$ refers to the (fixed) duration of each slot in a TDMA schedule. When a node is powered-on, it transits from $init$ state to \emph{listen} state and listens for the duration $\alpha_i^{listen}$ to determine if there is a synchronous set of nodes communicating on the medium. Similarly a node in \emph{coldstart} state waits for reception of frames until clock $x$ reaches the value of its timeout. If it receives such a frame, it enters the \emph{active} state, else it broadcasts another frame, loops into the coldstart state, and waits for another $\alpha_i^{cs}$ time units. For a brief description of the TTA startup algorithm, the reader is referred to~\cite{DS04}. For a detailed exposition to startup protocols, we refer the reader to~\cite{SP02}.

Denoting the minimum of all timeout values in any state by $y$, a timeout event in a state can be characterized by constraint $x = y$. Also the following properties might be of interest\footnote{For the formal semantic interpretation of these formulas see Section~\ref{tltl}.}.
\begin{itemize}
\item In each state, either a timeout occurs or it is set in the future:
$$\bx ((x = y) \vee (x < y))$$
\item If a node $i$ comes to the {\em listen} state (characterized by $p_{lis}$) at time $x = t_0$, it will move to the {\em coldstart} state (characterized by $p_{cs}$) in time no later than $x = t_0 + \alpha_i^{listen}$: $$\forall t_0.\bx (p_{lis} \wedge (x=t_0)\Rightarrow \dia (p_{cs} \wedge (x \leq t_0 + \alpha_i^{listen}))).$$
\end{itemize}

Clearly ordinary linear temporal logic (LTL)~\cite{P77,LP85} would not allow expressing such properties. One needs to extend LTL with the capability to express timeout constraints. Even the popular real-time extensions of LTL, e.g., TPTL~\cite{AH89} cannot be used in a straightforward manner to express these constraints. For example, in order to use TPTL for model checking the TTA start-up model discussed above, the model would have to be redesigned using explicit clock based frameworks e.g., timed automata~\cite{Alur94}. These clock based models in turn need to explicitly simulate the timeout semantics as discussed before. Also, as we shall discuss in Section~\ref{tltl-vs-tptl}, certain liveness properties on global timeout events cannot be expressed using TPTL.     

The primary objective of this work is to develop a real-time extension of LTL that can handle timeout contraints and possesses an efficient model checking algorithm as well. Over the past decade, there has been a sustained effort to increase the expressive power of temporal logic, which is a popular mechanism for specifying and verifying temporal properties of reactive and real-time systems. As we discuss further in Section~\ref{tltl-compare}, several attempts have been made to incorporate time explicitly into LTL, and to interpret the resulting logics over models that associate a time with every state. Examples of such logics are RTTL~\cite{O89}, XCTL~\cite{HLP90}, TPTL~\cite{AH89}, MTL~\cite{Koy90} etc. Quite a few verification tools have been developed based on these logics, {\em e.g.}, DT-SPIN~\cite{BD98a}, RT-SPIN~\cite{TC96}, UPPAAL~\cite{BDL04}. Since these tools adopt clock-based modeling approaches they can be used to formalize timeout systems only by first converting the timeout models into clock-based models (e.g., timed automata with clocks). On the other hand, the infinite bounded model checker of SAL (Symbolic Analysis Laboratory)~\cite{MOR+04} can model timeout systems, however supports only LTL model checking and demands considerable manual efforts while defining supporting lemmas and abstractions during the model checking process. In order to alleviate such problems timeout based modeling was earlier formalized by the authors in~\cite{ptd} in terms of predicate transition diagrams and the current work deals with defining the corresponding specification logic and model checking procedure.

The remainder of the paper is organized as follows: The logic TLTL is introduced  in the Section~\ref{tltl}. 
In Section~\ref{decide}, we introduce a monadic second order (MSO) theory of timeout state sequences and prove that TLTL when interpreted over discrete time can be embedded in it. 
In Section~\ref{tltl-compare}, we compare TLTL with other real-time extensions of LTL including XCTL, TPTL, and MTL. In Section~\ref{validity}, we describe a tableau based decision procedure for the validity (and satisfiability) checking of TLTL formulas followed by its complexity analysis. Model checking of TLTL formulas over timeout Kripke structures is discussed in Section~\ref{mc} with associated complexity analysis. In Section~\ref{undecide} we prove an undecidability result under dense time interpretation without time progress constraint. We conclude with a discussion on the directions for future work in Section~\ref{discussion}.

\section{The Logic TLTL}\label{tltl}

In this section we will define the syntax and semantics of {\it Timeout based Propositional Linear Temporal Logic}, TLTL.

\subsection{Syntax of TLTL}\label{tltl-syntax} The basic vocabulary of TLTL consists of a finite set $\cP$ of propositions $\true$, $\false$, $p, q, \ldots$, a finite set $T$ of (global) static timing variables $t_1, t_2, \ldots$ In addition, we allow a dynamic variable $x$ which represents the clock and a dummy variable $y$\footnote{Variable $y$ is essentially a place holder for minimum of the timeouts in a timeout program, which will be introduced in Section~\ref{timeout-program}. This abstraction is adopted primarily because, according to the behavior of a timeout system as discussed in Section~\ref{intro}, a discrete transition in a state may occur only when the current time is equal to the minimum valued timeout. For convenience, $y$ will also be referred sometimes as `minimum of the timeouts'.}. 
Assume $\Delta = \{<, =, >\}$, and let $\backsim$ range over $\Delta$. We use $\nR$ to denote the set of non negative real numbers, and $\N$ to denote the set of non negative integers.

\begin{itemize}
\item The set of atomic formulas ($\mathit{A}_f$) consists of propositions in $\cP$ and atomic constraints of the form $x < y, x = y, x < u, x = u,$ and $x > u$ where $u ::= t + c\;|\; c$, $t$ ranges over $T$ and $c \in \N$ is a constant. We will refer $x < u, x = u$ and $x > u$ as \emph{\sf {static constraints}} and $x < y$ and $x = y$ as \textit{\sf dynamic constraints}. 
\item (unquantified) Formulas are built using the following grammar \[ \phi ::= \mathit{a}_f \;|\; \phi \vee \phi \;|\; \neg \phi \;|\; \bigcirc \phi \;|\; \phi\ \unt \phi \] where $\mathit{a}_f$ ranges over $\mathit{A}_f$.
\item Finally, a quantified formula is built using universal quantification over timing variables at the outermost level as: $$\psi ::= \forall t_1 t_2\ldots t_k. \phi,$$ where $T_\psi = \{t_1, t_2, \ldots, t_k\} \subseteq T$ is the set of  timing variables appearing in the (unquantified) formula $\phi$.
\item The additional operators $\wedge, \Rightarrow, \Leftrightarrow$ and modal operators $\dia, \bx$ are introduced as abbreviations, $p \Rightarrow q \equiv \neg p \vee q, \dia \phi \equiv \true\ \unt \phi, \bx\phi\equiv \neg(\dia\neg\phi)$.
\end{itemize}

\subsection{Semantics of TLTL}\label{sem-tltl}
We consider the following point-wise  or (timeout) event based semantics for TLTL. Towards defining a model for a TLTL formula consider a sequence of states of the form $$\sigma: s_0 s_1 \ldots,$$ such that each $s_i$ gives a boolean interpretation ($\true, \false)$ to the propositions, and non negative real valued interpretation to the timing variables in $T$, to the clock variable $x$, and the variable $y$.

In a state $s_i$, let us assume that $s_i(x)$ denotes the value of the clock variable $x$, $s_i(y)$ the value of variable $y$, and $s_i(t_j)$ the value of timing variable $t_j \in T$. It is further required that 
\begin{description}
\item[$(m_1)$] \textit{Monotonicity:} Clock $x$ and variable $y$ do not decrease: $$\forall i:\ s_i(x) \leq s_{i+1}(x)\ \textit{and}\ \ s_i(y) \leq s_{i+1}(y)$$  
\item[$(m_2)$] \textit{Time Progress:} To ensure effective time progress in the model a \textit{divergence} condition\footnote{This is also known in the literature as `non-zenoness' or `finite-variability' condition.}, which says that time eventually increases, is required: $$\forall \delta \in \nR: \exists i \ \mbox{such that}\ s_i(x) > \delta$$
\item[$(m_3)$] \textit{State Transition:} Upon a change of state either timeout variables stay constant or some of them increase, that is, for each $i$: 
\begin{itemize} \item if the clock in state $s_i$ is less than the minimum of the timeouts, i.e. $y$, clock advances to this value in the next state $s_{i+1}$: $$[(s_i(x) < s_i(y)) \Rightarrow (s_{i+1}(y) = s_i(y)) \wedge (s_{i+1}(x) = s_{i}(y))]$$
\item else, if the clock in state $s_i$ is equal to the value of $y$, in the next state $s_{i+1}$, $y$ advances in the future: $$[(s_i(x) = s_i(y)) \Rightarrow (s_{i+1}(y) > s_i(y)) \wedge (s_{i+1}(x) = s_{i}(x))]$$
\end{itemize}
As a consequence we have for each $i$, $s_i(x) \leq s_i(y)$, that is, timeouts are always set in the future. \\
\item[$(m_4)$] \textit{Initiality:} For the initial state $s_0$ the following hold: either, $s_0(x) = s_0(y) = 0$ (when $x = y$ holds in $s_0$) or $0 = s_0(x) < s_0(y)$ (when $x < y$ holds in $s_0$). \\
\item[$(m_5)$] \textit{Constant Interpretation for Static Variables:} All the states are required to assign the same interpretation to the static timing variables, that is, for a given formula $\psi$, 
$$\forall t_j \in T_\psi: \, s_i(t_j) = s_0(t_j), \;\; \mbox{for each $i$}.$$
\end{description}
 
Thus a model for a TLTL formula may contain infinitely many  different states with different values of the clock and timeout variables. Boolean and modal operators are given the usual interpretation. We mention only atomic formulas in $\mathit{A}_f$ which are interpreted in a state as follows.
\[
\begin{array}{lll}
s_i \models p &\ \ \mathit{iff}\ \ & s_i(p) = \true\\
s_i \models x \backsim t_j + c &\ \ \mathit{iff}\ \ & s_i(x) \backsim s_i(t_j) + c\\
s_i \models x \backsim' y &\ \ \mathit{iff}\ \ & s_i(x) \backsim' s_i(y)\\
\end{array}
\]
Finally, we define $\sigma \models \psi \, \,  \mathit{iff}\, \,  s_0 \models \phi$ for any interpretation of the static timing variables appearing in $\phi$ {\em given in state $s_0$}. The formula $\psi$ is satisfiable (valid) if $\sigma \models \psi$ for some (all) sequence(s) $\sigma$. 

For example, consider \textsf{\small time bounded response property}, which specifies that ``event $q$ is always followed by event $p$ within $5$ time units''. It can be expressed by a TLTL formula \begin{equation}\label{eq0} \forall t_0. \bx(p \wedge (x = t_0) \Rightarrow \dia(q \wedge (x \leq t_0 + 5)))\end{equation}
We can also consider a variant of this as a \textsf{\small bounded timeout response property} stating that ``timeout event $q$ is always followed by timeout event $p$ within $5$ time units'', which would be expressed by a TLTL formula \begin{equation}\label{eq00} \forall t_0. \bx(p \wedge (x = y) \wedge (x = t_0) \Rightarrow \dia(q \wedge (x = y) \wedge (x \leq t_0 + 5)))\end{equation}
A quantified formula $\psi$ is termed as {\em closed} if all the timing variables appearing in it are bounded by a universal quantifier. In the rest of the discussion we will only consider closed quantified formulas. Also we will follow usual notational convention~\cite{O89,PH88,HLP90} of implicit universal quantification and would often drop the outermost universal quantification over global static timing variables in $T$. For example, the time bounded response property, specified by the TLTL formula (\ref{eq0}) would actually be written as 
\begin{equation*}\bx(p \wedge (x = t_0) \Rightarrow \dia(q \wedge (x \leq t_0 + 5))) \tag{$\phi_{BR_{TLTL}}$}\end{equation*} 
A formula of the form $x \leq z$ $(z:= u|y)$ is an abbreviation for $(x<z) \vee (x=z)$. Similarly $x \geq u$ abbreviates $(x>u) \vee (x=u)$. 
Note that $x > y$ is not a valid formula in TLTL. 

\section{An Embedding of TLTL in MSO}\label{decide}

In this section we explore the relationship of TLTL with monadic second order logic (MSO) with time. 
 Towards that we consider an interpretation of MSO in integer time structure. Subsequently we provide a straightforward meaning preserving translation between TLTL and monadic logic. 

\subsection{Monadic Second Order Theory of Timeout State Sequences}

Next we briefly recall the theory of timed state sequences $\cLT$ as introduced in~\cite{AH93} and extend it slightly. This is defined by adding a linearly ordered time domain $(TIME, \prec)$ with the theory of state sequences, S1S~\cite{Buchi60}, through a monotonically non decreasing function $f: \N \mapsto TIME$ that associates a time with every state in the sequence.
Thus a timed state sequence is a pair $(\sigma', f)$ consisting of an infinite sequence of states $\sigma' = s'_0s'_1\ldots$ and function $f$.

Let us additionally consider monotonically non decreasing function $g: \N \mapsto TIME$ representing the minimum of the timeouts in a state. This defines a {\it timeout state sequence} as a triple $(\sigma', f, g)$.

Let $\cLT$ be a second-order language with two sorts, a state sort and a time sort as considered in ~\cite{AH93}. The vocabulary of the congruence free sub language of $\cLT$ consists of:
\begin{itemize}
	\item The sets $\mathit{Var_{1}}$ and $\mathit{Var_{2}}$ for state sort. Set $\mathit{Var_{1}} = \{i, j, \ldots \}$ consists of individual (first order) variables and the set $\mathit{Var_{2}} = \{p, q,\ldots \}$ contains (second-order) set or predicate variables.
	\item The binary predicate symbol $<$ over the state and time sort;
	\item The unary function symbol $f$ from the state sort into the time sort;
	\item The quantification over individual variables in $\mathit{Var_{1}}$ and over predicate variables in $\mathit{Var_{2}}$.
\end{itemize}

Let $\cLTo$ be the language which in addition to $\cLT$ also contains:
\begin{itemize}
  \item The unary function symbols $g$ from the state sort into the time sort;
  \item The set of additional unary function symbols $\mathit{Var^t_{2}} = \{\mathbf{t_1, t_2},\ldots \}$ from the state sort into the time sort;
\end{itemize}
We consider only those formulas which do not contain any free individual variables. Further, we restrict our attention to structures that choose the set of natural numbers $\N$ as domain for both sorts with usual linear order $<$ on them. Given a formula $\phi$ of $\cLTo$ with the free predicate variables $p_1, \ldots, p_n \in \mathit{Var_{2}}$ and free function symbols $\mathbf{T_\psi} = \{\mathbf{t_1, \ldots, t_k}\} \subseteq \mathit{Var_2^t}$, an interpretation $I$ for $\phi$ specifies the sets $p_1^I, \ldots p_n^I \subseteq \N$, monotonically non decreasing  functions $f^I : \N \mapsto \N$ and $g^I : \N \mapsto \N$, $\mathbf{t}_1^I : \N \mapsto \N, \ldots,$ $\mathbf{t}_k^I : \N \mapsto \N$. The satisfaction relation $\models$ is defined in a standard fashion.

Every interpretation $I$ for $\phi$ implicitly defines a timeout state sequence $(\sigma', f, g)$: Let $\sigma'$ be the infinite sequence of states $s'_0 s'_1\ldots$, where $s'_i \in 2^{\{p_1, \ldots, p_n\}} \times \N^k$ such that 
$(p_j, (n_1, \ldots, n_k))  \in s'_i \Leftrightarrow i \in p_j^I$ and $\forall 1 \leq j \leq k. \mathbf{t_j}^I(i) = n_j$. Also let $f = f^I$ and $g = g^I$ for notational convenience. 

$\cLTo$-formulas define properties of timeout state sequences. For example, a bounded timeout response property discussed earlier (ref. Eq. (\ref{eq00})), and can be defined by a formula \begin{equation*}\forall i. (p(i) \wedge (f(i) = g(i)) \Rightarrow \exists j\geq i.(q(j) \wedge (f(j) = g(j)) \wedge (f(j) \leq f(i) + 5)))\tag{$\phi_{BR_{\LT}}$}\end{equation*}
An $\cLTo$-formula $\phi$ is satisfiable (valid) if it is satisfied by some (every) timeout state sequence.
The (second-order) theory of timeout state sequences is the set of all valid formulas of $\cLTo$. The following result is an immediate adaptation of the decidability result from~\cite{AH93}:

\begin{Fact}[Decidability] The validity problem for the language $\cLTo$ is decidable. \end{Fact}

\subsection{TLTL as a fragment of $\cLTo$}

Now we provide a meaning preserving compositional translation of TLTL formulas into $\cLTo$. Every TLTL-formula $\psi:= \forall t_1\ldots t_k . \phi$ can be translated into $\cLTo$, while preserving the set of models of $\psi$. The translation will use 
$\mathbf{T_\psi} = \{\mathbf{t_1, \ldots, t_k}\}$ to capture the static timing variables in $T_\psi = \{t_1, \ldots, t_k\}$, and a free individual variable $i \in  \mathit{Var_{1}}$ acting as a state counter. For every proposition $p$ of TPTL, we use a corresponding unary predicate $p(i)$ of state sort. We translate a TLTL-formula $\psi$ to the $\cLTo$-formula
\[\mathit{Tr(\psi)} = \forall i. \forall {\mathbf t_j} \in \mathbf{T_\psi}. \left[\Lambda_{m_2} \wedge \Lambda_{m_{3}} \wedge \Lambda_{m_4} \wedge \Lambda_{m_5} \wedge \mathit{Tr_0(\phi)}\right] \] where semantic constraints of TLTL as defined in Section~\ref{sem-tltl} are encoded by $\Lambda_{m_2} \ldots \Lambda_{m_5}$:
\[
\begin{array}{ll}
\Lambda_{m_2}: & \forall l \in \N. \exists m \in \N. f(m) > l \\
\Lambda_{m_3}: & [f(i) < g(i) \Rightarrow (g(i+1) = g(i))\wedge(f(i+1) = g(i))] \\ & \bigvee [f(i) = g(i) \Rightarrow (g(i+1) > g(i))\wedge(f(i+1) = f(i))] \\ & \bigvee \neg[f(i) > g(i)] \\
\Lambda_{m_4}: & [(f(0) = 0)\wedge(g(0)=0)]\vee[(f(0) \geq 0)\wedge(f(0) < g(0))]\\
\Lambda_{m_5}: & \left(\displaystyle\bigwedge_{1 \leq j \leq k}(\mathbf{t_j}(i) = \mathbf{t_j}(0))\right)
\end{array}
\]
The mapping $Tr_i$, for $i \geq 0$, is defined by induction on the structure of TLTL-formulas.
\[
\begin{array}{lll}
Tr_{i}(\false) & = & \false \\
Tr_{i}(p) & = & p(i) \\
Tr_{i}(x \backsim y) & = & f(i) \backsim g(i) \\
Tr_{i}(x \backsim t_j+c) & = & f(i) \backsim \mathbf{t_j}(0) + c\\
Tr_{i}(\phi \vee \varphi) & = & Tr_{i}(\phi) \vee Tr_{i}(\varphi) \\
Tr_{i}(\bigcirc \phi) & = & Tr_{i+1}(\phi)\\
Tr_{i}(\phi\ \unt \varphi) & = & \exs j \geq i. (Tr_{j}(\varphi) \wedge \for i \leq k < j. Tr_{k}(\phi))\\
\end{array}
\]
Given a model $\sigma = s_0,s_1, \ldots$ of TLTL-formula $\psi$, we can associate an $\cLTo$ interpretation $\cI = (\sigma', f, g)$ with $\mathit{Tr(\psi)}$ by making $p(i) = 1$ if $s_i \models p$ and $f(i) = s_i(x), g(i) = s_i(y)$, and $\mathbf{t_j}(i) = s_0(t_j)$. Similarly, given an $\cLTo$ interpretation $\cI = (\sigma', f, g)$ we generate a model $\sigma= s_0,s_1, \ldots$. Now by structural induction on $\phi$ we can prove the following:

\begin{Theorem}
Let $\psi$ be a TLTL formula. Then for a given model $\sigma$ of $\psi$, we have, $\sigma \models \psi$ if and only if $(\sigma', f, g) \models \mathit{Tr(\psi)}$.
\end{Theorem}

\section{A Comparison of TLTL with Other Logical Formalisms}\label{tltl-compare}

The most popular formalism for specifying properties of reactive systems is the linear temporal logic~\cite{P77,LP85}. The automatic verification and synthesis for finite state systems is usually carried out using the tableau-based satisfiability algorithm for a propositional version of the linear temporal logic (PLTL)~\cite{LP85}. PLTL is interpreted over models which retain only temporal ordering of the states by abstracting away the actual time instants at which events occur. However real-time systems call for explicitly expressing real-time constraints to reason about them, such as the \emph{bounded response} property which necessitates the development of formalisms which can express explicit time.

There are several approaches to extend LTL to express timing constraints. The first approach incorporates an explicit variable $x$, which expresses the current time without introducing any extra temporal operators. 
This is referred to as {\it explicit clock approach}, since the only new element introduced is the explicit clock variable. 
 An example of a first-order explicit clock logic is Real Time Temporal Logic (RTTL)~\cite{O89}, which is defined without restrictions on the assertion language for atomic timing constraints. A propositional version of this logic, called XCTL (Explicit Clock Temporal Logic), is discussed in~\cite{HLP90}. This logic allows integer variables to record the values of the global clock at different states, and integer expressions over these variables.

An alternative approach to express timing properties in a temporal logic has been to introduce a bounded version of the temporal operators. For example, a bounded operator $\dia_{[2,4]}$ is interpreted as ``eventually within $2$ to $4$ time units''. Using this notation we can write the time-bounded response property discussed earlier as: \begin{equation*} \bx(p \Rightarrow \dia_{[0,5]}q) \tag{$\phi_{BR_{MTL}}$}\end{equation*}
This approach for the specification of timing properties has been advocated by Koymans~\cite{Koy90} and is known as as Metric Temporal Logic (MTL). 

In yet another approach, time in a state is accessed through a quantifier, which binds (``freezes'') a variable to the corresponding time. This idea of freeze quantification was introduced by Alur and Henzinger in~\cite{AH89} in a logic known as TPTL (Timed Propositional Temporal Logic). The freeze quantifier ``$x.$'' binds the associated variable $x$ to the time of the current  temporal context; the formula $x.\phi(x)$ holds at time $t_0$ $\mathit{iff}$ $\phi(t_0)$ does. Thus in a formula $\dia x . \phi$, time variable $x$ is bound to the time of the state at which $\phi$ is ``eventually'' true.
By admitting atomic formulas that relate the time instants of different states, the time-bounded response property can be written as: \begin{equation*}\bx x_p.(p \Rightarrow \dia x_q. (q \wedge x_q \leq x_p + 5))\tag{$\phi_{BR_{TPTL}}$}\end{equation*}

\subsection{TLTL vs XCTL}\label{tltl-vs-xctl}
The logic XCTL as described in~\cite{HLP90}, contains static timing variables and an explicit clock variable in its vocabulary. An atomic formula $\mathit{a}_f$ is either an atomic proposition in $\cP$ or a constraint of the form $x \sim u$ or $c \sim u$, where $u = a_0 + a_1 \ast t_1 + \cdots a_m \ast t_m$ with constants $a_0, a_1 \ldots \in \N$ and $c \in \N$, and $t_0, t_1, \ldots, t_m$ being static timing variables.

XCTL formulas are built using the following grammar
\[ \phi ::= \mathit{a}_f \;|\; \phi \vee \phi \;|\; \neg \phi \;|\; \bigcirc \phi \;|\; \phi\ \unt \phi \] where $\mathit{a}_f$ ranges over $\mathit{A}_f$.

A model for XCTL consists of a sequence of states, \[ \sigma: s_0s_1 \ldots,\] such that each state $s_i$ gives a boolean interpretation to the propositions and an integer interpretation to the timing variables and to the clock variable $x$. Similar to TLTL, all static timing variables appearing in a XCTL formula assume the same valuation in all the states.

When compared to TLTL, it turns out that there exist properties involving the dynamic variable $y$, which cannot be expressed in XCTL. For example, for a timeout model of a real-time system the following property can be expressed in TLTL, - ``timeout occurs infinitely often'':  \begin{equation} \label{xctltptl2} \Box \dia(x=y) \end{equation}
The following sequence of states satisfies (\ref{xctltptl2}), \begin{equation}\label{xctltptl2_seq} \{0, 0\}, \{0, 3\}, \{3, 3\}, \{3, 5\}, \{5, 5\}, \ldots \end{equation}
In case of XCTL, only way to effectively characterize the state sequences satisfying (\ref{xctltptl2}) is by using constraints of the form $x \sim u$ or $u \sim c$. However, since static timing variables need to be given the same value in all the states in a state sequence, an equality of the form $x = u$ involving only static timing variables and constants in the r.h.s. expression $u$ can hold true only for a single value of $x$ (and $u$) in only finitely manly states in a state sequence, where $x$ assumes this value. Therefore, we need an infinite disjunction of such equalities to express (\ref{xctltptl2}) in XCTL, implying that there cannot exist any syntactically correct XCTL formula which can effectively characterize the state sequences similar to the one given in (\ref{xctltptl2_seq}) satisfying (\ref{xctltptl2}). 

On the other hand, consider XCTL formula \begin{equation} \label{xctltptl3} \Box(p \wedge (x = t_p) \Rightarrow \Box(q \wedge (x = t_q) \Rightarrow \Box(r \wedge (x = t_r) \Rightarrow [t_q - t_p \leq t_r - t_q]))) \end{equation}
This formula specifies that delay between events $p$ and $q$ is always less than the delay between $q$ and $r$. This property cannot be  specified in TLTL owing to the exclusion of the inequalities involving more than one static timing variable. 
 
\subsection{TLTL vs TPTL}\label{tltl-vs-tptl}
In~\cite{AH89}, Alur and Henzinger proposed an extension of LTL that is capable of relating the times of different states. For this purpose, they use {\it freeze quantification} by which every variable is bound to the time of a particular state. TPTL allows infinite number of variables $V = \{x_1, x_2, x_3, \ldots\}$ over which freeze quantification can be applied. The formulas of TPTL are built using the following grammar,
\[\phi::= \mathit{a}_f \,|\, \phi \vee \phi \;|\; \neg \phi \;|\; \bigcirc \phi \;|\; \phi\ \unt \phi \;|\; x_i.\phi \]
where $\mathit{a}_f$ is either an atomic proposition from $\cP$ or a constraint of the form $u_1 \leq u_2$, $u_1 \equiv_d u_2$, where $u_1,u_2: = x_i + c \, | \, c$ and $c \geq 0, d \geq 2$ are integer constants. Together they form the set of atomic formulas $A_f$. A variable $x_i$ can be bound by a freeze quantifier as ``$x_i.$'', which ``freezes'' $x_i$ to the time of local temporal context. Only closed formulas, where every occurrence of a variable is under the scope of a freeze quantifier, are considered.

The semantics for TPTL formulas is given by a sequence of states $\sigma = s_0,s_1, \ldots$ and an interpretation (environment) for the variables in $V$, $\cE : V \rightarrow \N$. The underlying time domain is taken to be the set of natural numbers $\N$. As before, each state assigns a Boolean interpretation to the propositions, and a (weakly) monotonic integer interpretation to a (hidden) global timing variable $\tau$, which is not used in the syntax of the formulas. 
We consider only atomic formulas in $A_f$ and formulas with freeze quantifiers. Let $\cE(x_i+c) = \cE(x_i) + c$ and $\cE(c) = c$. Also let $\cE[x_i:=a]$ denote the environment that agrees with the environment $\cE$ on all variables except $x_i$, and maps $x_i$ to $a \in \N$.
\[
\begin{array}{lll}
s_i \models_{\cE} p &\ \ \mathit{iff}\ \ & s_i(p) = \true\\
s_i \models_{\cE} u_1 \leq u_2 &\ \ \mathit{iff}\ \ & \cE(u_1) \leq \cE(u_2) \\
s_i \models_{\cE} u_1 \equiv_d u_2 &\ \ \mathit{iff}\ \ & \cE(u_1) \equiv_d \cE(u_2) \\
s_i \models_{\cE} x_i.\phi &\ \ \mathit{iff}\ \ & s_i \models \phi[\cE(x_i) = s_i(\tau)]
\end{array}
\]
A timed state sequence $\sigma$ is a model of a closed formula $\phi$ {\it iff} $s_0 \models_{\cE} \phi$ for any environment $\cE$.

As already noticed in~\cite{alursurvey91}, the static constraints in TLTL can play the same role as the freeze quantifier plays in TPTL. For example, consider the time-bounded response property $\phi_{BR_{TLTL}}$ again. This will be satisfied by only those models, which exactly assign a value to $t_0$, which is also the clock valuation at the instance of the occurrence of the event $p$, and therefore it is equivalent to the TPTL formula $\phi_{BR_{TPTL}}$. 
In general, assuming the same set of atomic constraints, a TPTL formula $$x.\phi$$ is equivalent to the TLTL formula \begin{equation}\label{eq-1j}\forall t_0. (x = t_0 \Rightarrow \phi)\end{equation}
However, this apparent syntactic correspondence is not without its problems. TPTL allows defining timing constraints referring to time instances of two past states e.g., $$\bx t_1.\bigcirc t_2.\dia (alarm \wedge t_2 > t_1 + 5)$$ This formula states that from now, if the time difference between two successive states is more than $5$ units, eventually an $alarm$ would be raised. 
Since TLTL does not allow referring to two past time instances, there is no syntactically straightforward translation for such formulas in TLTL using (\ref{eq-1j}) above. 
However as it turns out, this is really not a problem because such formulas involving reference to two past timing instances are semantically equivalent to formulas which demand referring to only one previous time instance in the state, where the second timing instance would be frozen. In this example an semantically equivalent TPTL formula would be $$\bx t_1 .\bigcirc t_2.(t_2 > t_1 + 5 \Rightarrow \dia (alarm))$$ which can be translated into an equivalent TLTL formula using (\ref{eq-1j}) (omitting the outermost quantification) $$\bx (x = t_1 \Rightarrow \bigcirc (x = t_2 \Rightarrow ((x > t_1 + 5) \Rightarrow \dia (alarm))))$$
It may be also noted that TLTL is suitable in case where one needs to express formulas about timed systems with timeouts as the following kinds of condition cannot be expressed in TPTL - ``timeout always occurs in the next state of time increment.'' \begin{equation}\label{eq-0}\Box((x < y) \Rightarrow \bigcirc (x = y))\end{equation} 
The reason that there cannot be any formula in TPTL, which can characterize exactly the same set of models as the formula (\ref{eq-0}) does is as follows.  Since $x$ refers to the time(s) when (\ref{eq-0}) holds, these can only be captured using freeze quantifier in TPTL. Now since TPTL inequalities only involve (frozen) variables or constants, for variable $y$ also, we need to use these. However, $y$ being a dynamic variable would assume infinitely many different values in a model of the formula (\ref{eq-0}), these values cannot be captured using constants (or else would demand infinitely many constant based inequalities of the form $[x < c \Rightarrow \bigc(x = c)]$). Therefore, the only option is to potentially use variables under freeze quantifier. However, the inequality $x < y$ would demand that such variable (that is $y$) must refer to a future state, since time flows only in the forward direction, in particular, the next state itself. A formula like the one below may (appear to) capture such a scenario. \begin{equation}\label{eq-1}\Box x.((x < y) \Rightarrow \bigcirc y.(x = y))\end{equation} However atomic constraints in TPTL cannot refer to the time points of the future states as is evident from the very syntax of the freeze quantifier, e.g., in case of the TPTL formula (\ref{eq-1}). First, $y$ is a free variable and then $y$ is bound by the (second) freeze quantifier and therefore, both $y$s are actually different variables - such formulas involving free variables are in any case not allowed in TPTL. Thus, neither constants nor timing variables based inequalities can be used to express the inequalities appearing in the formula (\ref{eq-0}). That is why, the state sequences satisfying TLTL formula (\ref{eq-0}) cannot be characterized in TPTL. 
 
On the other hand, there are formulas in TPTL, which cannot be characterized in TLTL. For example, consider the state sequences, in which ``an event $p$ occurs at all even time points.'' This can be characterized by the TPTL formula $\Box x.(x \equiv_2 0 \Rightarrow p)$. However, as proved in~\cite{AH93}, this property is not expressible without congruences. This in turn, implies that due to the nature of arithmetical constraints, this TPTL formula cannot be expressed in TLTL.

\subsection{TLTL vs MTL}

MTL~\cite{Koy90} extends LTL by constraining the temporal operators on (bounded or unbounded) intervals of the real numbers specified as subscripts. The formulas in MTL are inductively built using the following grammar
\[\phi::= p \,|\, \phi \vee \phi \;|\; \neg \phi \;|\; \phi\ \unt_{I}\ \phi \]
where $p \in \cP$ is a proposition and $I$ is a (bounded or unbounded) interval with integer (or rational) end-points.  An \textit{interval} is a nonempty convex subset of $\nR$, which may assume one of the following forms: $[a,b]$, $[a,b)$, $[a,\infty)$, $(a,b]$, $(a,b)$, $(a,\infty)$, where $a \leq b$ for $a,b \in \nR$. The interval $I$ is singular \textit{iff} it is of the form $[a,a]$ (also written as $=a$).

The formulas of MTL can be interpreted over a timed state sequence $(\sigma, f)$, where $\sigma = s'_0,s'_1,\ldots$ is a untimed state sequence giving Boolean interpretation to the propositions and $f: \N \mapsto \nR$ is a mapping such that $f(i)$ denotes the time at state $s'_i$. 
The satisfaction relation $(\sigma, f)\models \phi$ is defined in a usual way. We only mention the case of the formula $\phi\ \unt_{I} \varphi$: 
\begin{multline} \nonumber
(s_i, f(i)) \models \phi\ \unt_{I}\ \varphi \ \ \mathit{iff}\ \ \exs j \geq i. [((s_{j},f(j)) \models \varphi) \\ \bigwedge (\for i \leq k < j. (s_{k},f(k)) \models \phi) \wedge (f(j) \in f(i) + I)], 
\end{multline}
where $f(i) + I$ is defined using simple rules of interval arithmetic, e.g., if $I = [a,b]$, then $f(i) + I$ stands for the interval $[f(i)+a, f(i)+b]$.

Since it is well known that the satisfiability and model-checking problems for MTL are undecidable over the state-based semantics (under $\nR$), we will consider a fragment of MTL known as {\it Metric Interval Temporal Logic} (MITL) introduced by Alur et al.~\cite{alur1996benefits}, in which the temporal operators can only be constrained by nonsingular intervals. Thus `punctuality properties' like $\dia_{=3}p$ (``eventually exactly after 3 time units $p$ would hold'') cannot be specified in MITL.

It is known that any MITL formula can also be expressed in TPTL~\cite{AH93}. Specifically, if the atomic constraints permit comparison and addition of constants, then MITL formula \begin{equation}\label{jjj} \phi\ \unt_{I} \varphi \end{equation} is equivalent to the TPTL formula $$x. \phi\ \unt z.(\varphi \wedge z \in x+I)$$ where $z \in x+I$ can be expressed using TPTL constraints given the boundaries of $I$. It has been shown recently in~\cite{BCM05} that TPTL is strictly more expressive than MTL for both point-wise and interval-based semantics. Now in the light of the discussion presented in previous Section~\ref{tltl-vs-tptl}, it is easy to see that any MITL formula can also be expressed in TLTL. Specifically, MITL formula (\ref{jjj}) is equivalent to the TLTL formula $$\forall t_0. (x = t_0 \Rightarrow \phi\ \unt (\varphi \wedge x \in t_0+I)$$ where $x \in t_0+I$ can be expressed using atomic constraints in TLTL, given the boundaries of $I$. For example, MITL formula, $\Box( p \Rightarrow \dia_{[2, 5]} \, q)$ can be expressed in TLTL as $$\forall t_0. \Box (p \wedge x = t_0 \Rightarrow \dia (q \wedge x \geq t_0 + 2 \wedge x \leq t_0 +5 )).$$ Also, on the other hand, there exist TLTL formulas (e.g., one given in (\ref{eq-0})), which cannot be expressed in MTL under point-wise semantics.

\section{Decision Procedure for Validity of TLTL formulas}\label{validity}

We consider a decision procedure for checking the validity of TLTL formulas employing similar techniques used in~\cite{HLP90}. In order to check the validity of a given TLTL formula $\psi=\forall t_1\ldots t_k. \phi$, we take the negated formula $\neg\phi$ and actually check for its satisfiability using a tableau like construction by posing the question, `are there positive real values for the timing variables $t_1, \ldots, t_k$ that will make the formula $\neg\phi$ satisfiable?' 

\subsection{Closure of a Formula}

Let $\phi$ be a TLTL formula, which is to be checked for satisfiability. We define the Fischer-Ladner closure $Cl(\phi)$ as the least set containing $\phi$ and closed under the following:
\begin{description}
\item[$(c_1)$] $\true,\ \false, \bigcirc \true \in Cl( \phi)$,
\item[$(c_2)$] $\forall p \in {\cal P}_{\phi}$, $p, \neg p \in Cl(\phi)$, where ${\cal P}_{\phi}$ is
the set of atomic propositions appearing in $\phi$,
\item[$(c_3)$] $\neg \psi \in Cl(\phi) \Leftrightarrow \psi \in Cl(\phi)$ -- we identify $\neg \neg
\psi$ with $\psi$ and $\neg \true$ with $\false$,
\item[$(c_4)$] $\psi \vee \psi' \in Cl(\phi) \Rightarrow \psi,\ \psi'\ \in Cl(\phi)$,
\item[$(c_5)$] $\bigcirc \psi \in Cl(\phi) \Rightarrow \psi \in Cl(\phi)$,
\item[$(c_6)$] $\neg\bigcirc \psi \in Cl(\phi) \Rightarrow \bigcirc\neg\psi \in Cl(\phi)$,
\item[$(c_7)$] $\psi\ \unt \psi' \in Cl(\phi) \Rightarrow \psi,\ \psi',\ \bigcirc (\psi\ \unt \psi') \in Cl(\phi)$,
\item[$(c_8)$] $x \backsim y \in Cl(\phi) \Rightarrow x < y, x = y \in Cl(\phi)$,
\item[$(c_9)$] $x \backsim u \in Cl(\phi) \Rightarrow x \backsim' u \in Cl(\phi)$ for every $\backsim' \in \Delta$,
\item[$(c_{10})$] $x < y \in Cl(\phi) \Rightarrow \bigcirc (x = y), \dia(x<y) \in Cl(\phi)$
\item[$(c_{11})$] $x = y \in Cl(\phi) \Rightarrow \bigcirc (x < y), \dia(x=y) \in Cl(\phi)$
\item[$(c_{12})$] $x \backsim u \in Cl(\phi) \Rightarrow \dia(x>u) \in Cl(\phi)$. 
\end{description}
Intuitively $Cl(\phi)$ includes all the formulae that play some role in deciding the satisfiability of $\phi$. Using structural induction on $\phi$, it can be shown that $|Cl(\phi)|$ $\leq 7|\phi| + 3$.

\subsection{Atoms}\label{atoms}
An atom $A \subseteq Cl(\phi)$ is a 
consistent set of formulas  such that
\begin{description}
\item[$(a_{1})$] $\true, \bigcirc \true \in A$.
\item[$(a_{2})$] For every $\psi \in A \Leftrightarrow \neg \psi \not \in A$.
\item[$(a_{3})$] For every $\psi \vee \psi' \in A \Leftrightarrow \psi \in A \,\, \mbox{or}\,
\, \psi' \in A$.
\item[$(a_{4})$] For every $\psi\ \unt \psi' \in A \Leftrightarrow \psi' \in A$ or $\psi, \bigcirc
(\psi\ \unt \psi') \in A$.
\item[$(a_{5})$] For every $x < y, x =y  \in Cl(\phi)$, precisely one of
them is in $A$.
\item[$(a_{6.1})$] For every $x < y \in A \Rightarrow \next{(x = y)} \in A$.
\item[$(a_{6.2})$] For every $x = y \in A \Rightarrow \next{(x < y)} \in A$.
\item[$(a_{7})$] For every $x \backsim u \in Cl(\phi)$, exactly one of $x<u, x= u, \,\,\mbox{or}\,\, x>u$ is in $A$.
\item[$(a_{8})$] If $C(A)$ denotes the set of all constraints in $A$, it is required that $C(A)$ forms a consistent set. In particular, for every $x \backsim u_i \in Cl(\phi)$, $x \backsim u_i \in C(A)$ only if exactly one of the following holds, where we let $H_A = \bigwedge_{x \backsim u_j \in C(A)} (x \backsim u_j)$:

$\left\{ \begin{array}{ll} & x < y \in A \mbox{ and } (x < y)\wedge (x \backsim u_i)\wedge H_A \mbox{ is satisfiable over } \nR \ \ \mbox{\small \sf OR } \\
&  x = y \in A \mbox{ and } (x = y)\wedge (x \backsim u_i) \wedge H_A \mbox{ is satisfiable over } \nR \ \ \mbox{\small \sf OR }\\
&  x < y, x =y \notin A \mbox{ and } (x \backsim u_i) \wedge H_A \mbox{ is satisfiable over } \nR \\
\end{array} \right.$

Informally, we include a static constraint $x\backsim u_i \in Cl(\phi)$ in atom $A$ only if the resultant set of constraints in $A$ remains consistent.
\item[$(a_{9})$] For every $x \backsim u \in A \Rightarrow \true\ \unt (x > u) \in A$.
\end{description}
The requirement that every atom contains the formula $\bigcirc\true$ is to ensure that only infinite sequences will be considered as possible models. 

Additionally, we define two special atoms.
\begin{align*} 
A_{0_=} & = \{\true, \bigcirc\true, x = 0, x = y, \bigcirc(x < y), \true\unt{(x>0)}, &\\ 
& \hspace*{2.5cm} \bigcirc(\true\unt{(x>0)})\}, \mbox{ and} & \\ 
A_{0_<} & = \{\true, \bigcirc{\true}, x = 0, x < y, \bigcirc(x=y), \true\unt{(x>0)}, & \\
& \hspace*{2.5cm} \bigcirc(\true\unt{(x>0)}), \bigcirc(x>0) \}. &
\end{align*}
We denote the set of all atoms by $At$, which also contains $A_{0_=}$ and $A_{0_<}$.

\subsection{Tableau Construction}\label{tab-construct}
We construct a structure $\cA_\phi = (At, R)$, which is a directed graph with atoms as nodes; and its edges are defined by the relation $R$ as follows: 
\[
(A, B) \in R \Leftrightarrow \left\{ \begin{array}{ll} 1.\, &
\mbox{for every}\, \next a \in Cl(\phi), \\
& \next a \in A \Leftrightarrow a \in B, \mbox{where } a \in {\cal P}_{\phi} \cup C(\phi); \\
2.\, & \mbox{for every}\,\, x = u \in Cl(\phi), \\
& x = u \in A \Rightarrow  x = u \in B \,\, \mbox{or}\,\, x > u \in B;\\
3.\, & \mbox{for every}\,\, x > u \in Cl(\phi), \\
& x > u \in A \Rightarrow  x > u \in B;
\end{array} \right.
\]
where $C(\phi)$ refers to the set of atomic constraints appearing in $\phi$.

It is not difficult to see that under the definition of $R$, the following facts hold. 

\begin{Fact}\label{f1}
There is no atom $A \in At$ such that $(A, A_{0_=}) \in R$. 
\end{Fact}
\begin{Fact}\label{f2}
There is no atom $A \in At\setminus\{A_{0_=}\}$ such that $(A, A_{0_<}) \in R$. 
\end{Fact}

In other words, atom $A_{0_=}$ has no incoming edges and the only permissible incoming edge to atom $A_{0_<}$ is $(A_{0_=}, A_{0_<}) \in R$. $A_{0_=}$ and $A_{0_<}$ will be referred from now on as {\it initial atoms}. Also note that only states, where atom $A_{0_=}$ may hold are those which interpret both clock variable $x$ and minimum of the timeout variable $y$ as $0$.

Let $\cA'= (W',R')$ be a substructure of $\cA_\phi$ and let $\cC$ be a strongly connected subgraph (SCS) of $\cA'$.
\begin{itemize}
\item $\cC$ is said to be {\it terminal} in $\cA'$ if it has no outgoing edges.
\item $\cC$ is said to be {\em self-fulfilling} if every atom has a successor in $\cC$, and $\textit{for every }p\ \unt q \in A \in \cC, \textit{ there exists } B \in \cC\; \textit{such that} \; q \in B$.
\item $\cC$ is said to be {\em useless} in $\cA'$ if it is terminal in $\cA'$ but is not self-fulfilling.
\end{itemize}
\subsection{The Timing Relation between Atoms}

\subsubsection*{Relation between Successive Atoms:}

Consider two atoms $A,B$ from $\cA_\phi$ such that $(A,B) \in R$. Assume the set of constraints in $A$ to be $C(A) = T(A) \cup S(A)$, where $T(A) = \{T_{out}\}$ contains the (unique) dynamic constraint and $S(A)= \{S_1, \ldots, S_m\}$ the set of static constraints. Further, the set of constraints in $B$ is $C(B) = T(B) \cup S(B)$ where $T(B)\ =\ \{T_{out}'\}$, and $S(B)\ =\ \{S'_1, \ldots, S'_m\}$. For every $S_i$ there is a corresponding $S'_i$ and for $T_{out}$ there is a corresponding $T_{out}'$ such that:
\begin{itemize}
\item if $S_i$ is $x<u$, $S'_i$ is $x \backsim u$. This follows from the condition $(a_7)$ in Section~\ref{atoms} for defining an atom,
\item if $S_i$ is $x = u$, $S'_i$ is either $x = u$ or $x > u$. This follows from the condition $(2)$ for defining $R$ in Section~\ref{tab-construct},
\item if $S_i$ is $x > u$, $S'_i$ is also $x>u$. This follows from the condition $(3)$ for defining $R$ in Section~\ref{tab-construct}, and
\item if $T_{out}$ is $x < y$, $T_{out}'$ is $x = y$. Else if, $T_{out}$ is $x = y$, $T_{out}'$ is $x < y$. This follows from conditions $(a_{6.1}), (a_{6.2})$ in Section~\ref{atoms} and condition $(1)$ for defining $R$ in Section~\ref{tab-construct}.
\end{itemize}

The temporal relation between two atoms produces the following results, which allow us to select values for $x, y$ satisfying constraints in one atom, once the values for which these variables satisfy other constraints are known. Let us assume that $\chi, \chi'$ denote valuations for clock $x$, $\psi, \psi'$ for $y$, and $\alpha_1, \alpha_2, \ldots, \alpha_k$ for timing variables $t_1, t_2, \ldots, t_k$.

\begin{Lemma}\label{lemma1}
If $\chi', \psi', \alpha_1, \alpha_2, \ldots, \alpha_k$ are non negative reals satisfying $C(B)$, there exist non negative reals $\chi, \psi$ such that $\chi, \psi, \alpha_1, \alpha_2, \ldots, \alpha_k$ satisfy $C(A)$ and $\chi \leq \chi', \psi \leq \psi'$.
\end{Lemma}
{\bf Proof.} Assume $\chi', \psi', \bar{\alpha}$ satisfy $C(B)$, where $\bar{\alpha} = \alpha_1, \alpha_2, \ldots, \alpha_k$. We need to show that there exist $\chi \leq \chi',\, \psi \leq \psi'$ such that $\chi,\psi, \bar{\alpha}$ satisfy $C(A)$. We consider different cases.\\\\
{\sf Case 0:} If $C(A) = \emptyset$, that is, $\phi$ is a purely qualitative formula not involving any of static or dynamic constraints, choose $\chi = \chi'$ and $\psi = \psi'$. \\\\
{\sf Case 1:} If $S(A) = \emptyset$ but $T(A) \neq \emptyset$. We choose $\chi, \psi$ based upon the nature of $T_{out}$.
\begin{itemize}
    \item Let $T_{out} \equiv x = y \  \in C(A)$. 
    Now by the definition of timing relation between atoms $A$ and $B$, we have $T_{out}' \equiv x < y$ implying that $\chi' < \psi'$. So, choose $\psi = \chi = \chi'$.
    \item Let $T_{out} \equiv x < y \  \in C(A)$.
    Again by the definition of timing relation between atoms $A$ and $B$, we have $T_{out}' \equiv x = y \
    \in C(B)$. Therefore $\chi' = \psi'$. We choose $\psi = \psi'$ and some arbitrary value $\chi \in [0, \chi')$. Note that this is always feasible since in the only exceptional case when $\chi' = \psi' = 0$, $B$ would be an initial atom $A_{0_=}$ and thus $A$ cannot be present (see Fact~\ref{f1}).
\end{itemize}
{\sf Case 2:} $S(A) \neq \emptyset$ and there exists a constraint $S_i \in S(A)$ of the form $x = t_i + c_i$ or $x = c$ ($c_i, c$ are constants) as the case may be, then choose $\chi$ as $\alpha_i + c_i$ or $c$ which would necessarily satisfy all of $S_1, \ldots, S_m$ following the definition of an atom (condition ($a_8$)). Now based upon the nature of $T_{out}$, we will choose $\psi$ and prove the consistency of the choice.
\begin{itemize}
    \item Let $T_{out} \equiv x = y \  \in C(A)$. Choose $\psi = \chi$. Now by the definition of timing
    relation between atoms $A$ and $B$, we have $T_{out}' \equiv x < y$. Therefore $\chi' < \psi'$. Now $(x = t_i + c_i)\ \in S(A) \Rightarrow (x = t_i + c_i)$ or $(x > t_i + c_i) \ \in S(B)$ implying $\chi' \geq \alpha_i + c_i$. Thus we
    have, $\psi = \chi = \alpha_i + c_i \leq \chi' < \psi'$.
    Similarly, for $(x = c) \in S(A)$.
    \item Let $T_{out} \equiv x < y \  \in C(A)$. Choose $\psi$ such that $\alpha_i < \psi \leq \psi'$.
    Again by the definition of timing relation between atoms $A$ and $B$, we have $T_{out}' \equiv x = y \
    \in C(B)$. Therefore $\chi' = \psi'$. Also $(x = t_i + c_i)\ \in C(A) \Rightarrow (x = t_i + c_i)$ or $(x > t_i + c_i)
    \ \in C(B)$, which also means $\chi' \geq \alpha_i + c_i$, i.e., $\chi' \geq \chi$. Similarly, for $(x = c) \in S(A)$.
    \item $T(A) = \emptyset$. Choose $\psi = \chi$.
\end{itemize}
So, in all the situations we can choose $\chi$ and $\psi$ such that $\chi \leq \chi'$ and $\psi \leq \psi'$.\\\\
{\sf Case 3:} $S(A) \neq \emptyset$ and there does not exist any
constraint $S_i \in S(A)$ of the form $x = t_i + c_i$ or $x = c$. Let
\begin{itemize}
\item $E_l = \{\alpha_j + c_j  \mid (x > t_j + c_j) \in C(A)\} \cup \{c \mid (x > c) \in C(A)\}$ and $l =
\max(E_l)$ if $E_l \neq \emptyset$ else $l = -\infty$,
\item $E_m = \{\alpha_j + c_j \mid (x < t_j + c_j) \in C(A)\} \cup \{c \mid (x < c) \in C(A)\}$ and $m =
\min(E_m)$ if $E_m \neq \emptyset$ else $m = \infty$.
\end{itemize}
Note that $l < m$ since $\bigwedge_i S_i$ is satisfiable. Again, by the definition of timing relation between atoms $A$ and $B$, we have
$\forall w \in E_l.\ (x > w) \in S(A) \Rightarrow (x > w) \in S(B)$ implying that $l < \chi'$. Therefore, choose $\chi$ such that
\begin{equation}\label{eq1}
\begin{split}
& l < \chi \leq \chi'\  \mbox{if}\  \chi' < m \\
& l < \chi < m\  \mbox{if}\  \chi' \geq m
\end{split}
\end{equation}
Such a choice of $\chi$ satisfies all of $S_1, \ldots, S_m$. Now based upon the nature of $T_{out}$, we choose the values of $\chi$ and $\psi$ and prove the consistency of such a choice.
\begin{itemize}
    \item Let $T_{out} \equiv x = y \  \in C(A)$. Choose any value for $\chi$ satisfying (\ref{eq1}) and choose $\psi = \chi$. Since $\chi \leq \chi' < \psi'$, we have $\chi \leq \chi'$ and $\psi < \psi'$.
    \item Let $T_{out} \equiv x < y \  \in C(A)$. Choose $\psi = \chi'$. Because $T_{out} \wedge \bigwedge_i S_i$ is
    satisfiable, we must be able to choose $\chi$ such that $\chi < \psi$, which implies that $\chi < \chi'$.
    \item $T(A) = \emptyset$. Choose any value for $\chi$ satisfying (\ref{eq1}) then choose $\psi = \chi$.
\end{itemize}
So in both the situations we can choose $\chi$ and $\psi$ such that $\chi \leq \chi'$ and $\psi \leq \psi'$. 

Hence. \hfill \bbox

\subsubsection*{Relation between Atoms in a Self-Fulfilling SCS:}

In a self-fulfilling SCS every two atoms have the same set of static constraints, but they differ in the dynamic constraint.

\begin{Lemma}\label{lemma2}
Let $A$ and $B$ be two atoms in some self-fulfilling SCS $\cC$, then $S(A) = S(B)$, and all the static constraints must be of the form $x > u$.
\end{Lemma}
{\bf Proof.} Since $A, B \in \cC$ and $\cC$ is a SCS, hence by definitions of atom and relation $R$, $x>u \in S(A) \Leftrightarrow x>u \in S(B)$. It remains to show that $(x \backsim' u) \not \in S(A)$, where $\backsim' \in \{<,=\}$. Assume that it is not the case, which means, $x \backsim' u \in C(A)$. By the definition of an atom, $\true\ \unt(x>u) \in A$. Since $\cC$ is a self-fulfilling SCS, there
must be an atom $D \in \cC$ such that $(x>u) \in S(D)$. It follows that $(x>u) \in S(A)$ as well because $A$ is reachable from $D$, a fact that contradicts the definition of an atom. Therefore, we conclude that $x \backsim' u \not\in S(A)$. Since this will be true for atom $B$ as well, it follows $S(A) = S(B)$. \hfill \bbox

\begin{Lemma}\label{lemma3}
If $\chi, \psi, \bar{\alpha}$ is a satisfying solution for $C(A)$ and $A \in \cC$ (a self-fulfilling SCS), for every $B \in \cC$ such that $(A, B) \in R$, there exist $\chi', \psi'$, such that $\chi', \psi', \bar{\alpha}$ satisfy $C(B)$ and $\chi' \geq
\chi, \psi' \geq \psi$.
\end{Lemma}
{\bf Proof.} We consider only dynamic constraints appearing in $A$ and $B$:
\begin{itemize}
    \item $ (x = y) \  \in C(A) \wedge (x < y) \  \in C(B)$. Choose $\chi' = \chi$ and any $\psi' > \psi$.
    \item $ (x < y) \  \in C(A) \wedge (x = y) \  \in C(B)$. Choose $\chi' = \psi' = \psi$. 
    \item $T(A) = T(B) = \emptyset$. Choose arbitrarily $\chi', \psi' \in \nR$ such that $\chi' > \chi, \psi' > \psi,$ and $\chi' \leq \psi'$.
\end{itemize}
Note that from Lemma $2$, every atom in $\cC$ contains all other constraints of the same form $x > u$, which are immediately satisfiable
by any $\chi' \geq \chi$. \hfill \bbox

\subsection{Fulfilling Paths and Satisfiability}\label{sat-algo} 
An infinite path $\pi = A_0, A_1, \cdots,$ (where $A_0, A_1, \cdots$ are atoms) is called a {\it fulfilling path} for $\phi$ if for every $i \geq 0$:
\begin{enumerate}
\item $\phi \in A_0$.
\item  $(A_i, A_{i+1}) \in R$.
\item  For every $p\ \unt q \in Cl(\phi)$, if $p\ \unt q \in A_i$, then there exists some $j \geq i$ such that $q \in A_j$.
\end{enumerate}

\begin{Theorem} \label{theoremsat}
The formula $\phi$ is satisfiable if and only if there exists a fulfilling path for $\phi$ in $\cA_\phi$.
\end{Theorem}

{\bf Proof.} If $\phi$ is satisfiable and $\sigma$ is a model for it then the corresponding fulfilling path can be given by
$\pi = A_0, A_1, \cdots$, where $A_i=\{p \in Cl(\phi)\ \mid\ \sigma^i \models p\}$.

On the other hand let $\pi = A_0, A_1, \cdots, $ be a fulfilling path for $\phi$. Define a model $\sigma = s_0, s_1, \cdots, $ for $\phi$ such that each state $s_i$ ($\forall i \geq 0$), interprets proposition $p$ as ${\tt true}$ $\mathit{iff}$ $p$ 
is in $A_i$. Since $\pi$ is an infinite path, beyond a certain point (say $A_k$), all the atoms in $\pi$ must be repeating infinitely often. These infinitely repeating atoms must be reachable from each other, and hence must be contained in a self-fulfilling SCS $\cC$. Let $\alpha_1, \alpha_2, \ldots, \alpha_k, s_{k+1}(x), s_{k+1}(y)$ be any solution that satisfies $C(A_{k+1})$. Using Lemma~\ref{lemma1}, we can trace the path $\pi$ backwards till $A_0$ assigning values $(s_0(x) \leq s_{1}(x) \ldots \leq s_{k-1}(x) \leq s_{k}(x) \leq s_{k+1}(x), s_0(y) \leq s_1(y) \leq \ldots \leq s_{k -1}(y) \leq s_k(y) \leq s_{k+1}(y))$ to $(x, y)$ in atoms $A_0, A_1, \ldots A_k$ on the way, which satisfy constraints in $C(A_0), C(A_1), \ldots C(A_k)$. Also using Lemmas~\ref{lemma2},\ref{lemma3} we can assign values $( s_{k+1}(x)\leq s_{k+2}(x) \leq s_{k+3}(x) \ldots,\ \   s_{k+1}(y) \leq s_{k+2}(y) \leq s_{k+3}(y) \ldots)$ for the future states $s_{k+2}, s_{k+3}, \cdots,\ $. Clearly $\sigma$ is a infinite sequence of states satisfying the formula $\phi$. \hfill \bbox

From this theorem we conclude that it is sufficient to look for a fulfilling path for $\phi$ in $\cA_\phi$ in order to determine the satisfiability of $\phi$.

\subsection{Satisfiability Checking}
The fulfilling path for a TLTL formula $\phi$ can be constructed as follows:\\
\\
{\bf let} $\cA^{*} = (\cW^*, \cR^*) = \cA_\phi$ be the initial structure resulting from the construction described in the Section~\ref{tab-construct}.\\
{\bf while}($\cA^* \neq \emptyset$ OR $\cA^*$ does not contain any useless maximal SCS)\\
\hspace*{.6cm} {\bf begin}\\
\hspace*{1cm} {\bf let} $\cC$ be a useless maximal SCS in $\cA^*$\\
\hspace*{1.5cm} $\cW^* = \cW^* \setminus {\cC}$\\
\hspace*{1.5cm} $\cR^* = \cR^* \cap (\cW^* \times \cW^*)$\\
\hspace*{.6cm} {\bf end}\\
{\bf if} (there is an atom $A$ in $\cW^*$ such that $\phi \in A$)\\
\hspace*{1.5cm} {\bf then} report {\tt success} \\
{\bf else} report {\tt failure}.

\begin{Theorem}
The formula $\phi$ is satisfiable if and only if the above algorithm reports success.
\end{Theorem}

The algorithm succeeds if and only if the tableau $\cA_{\phi}$
contains a finite path $\pi = A_0, \ldots, A_k$ that starts at an
atom $A_0$, containing $\phi$, and reaches $A_k$ at a terminal
self-fulfilling SCS $\cC$. This path can be used to construct a
fulfilling path for $\phi$. Hence by Theorem~\ref{theoremsat},
$\phi$ is satisfiable if and only if the algorithm above reports
success.\hfill \bbox

\subsection{Complexity Analysis}\label{sat-complexity}

For the complexity analysis we would require the following result.

\begin{Lemma}\label{lemmac}
Checking that the constraints appearing in an atom are satisfiable over $\nR$ can
be done in time $O(|Cl(\phi)|)$.
\end{Lemma}
{\bf Proof.} There exists a well known polynomial time procedure~\cite{pratt77} to decide the satisfiability of a conjunction of
linear inequalities of the form $\xi \leq \eta + c$, where $\xi, \eta$ are real-valued variables and $c$
is an integer constant, by reducing the problem to the problem of deciding the nonexistence of a cycle with negative weight
in a weighted directed graph such that inequality $\xi \leq \eta + c$ induces two nodes corresponding to variables $\xi, \eta$
and an edge $(\xi, \eta)$ labeled with $-c$.

Nonetheless, owing to special nature of the constraints considered
here, we can show that a linear time procedure exists to check the
satisfiability of the constraints appearing in an atom. Let us partition
the the set of constraints appearing in atom $A$ as follows:
\[ C(A)= C_{xy} \cup C_{=c} \cup C_{=v} \cup C_{>c} \cup C_{>v}
\cup C_{<c} \cup C_{<v},\ \mbox{where}\]
\indent $C_{xy}$ consists of constraints of the form  $(x \backsim y)$,\\
\indent $C_{=c}$ consists of constraints of the form  $(x = c)$, \\
\indent $C_{=v}$ consists of constraints of the form $(x = t + c')$, \\
\indent $C_{>c}$ consists of constraints of the form $(x > c)$, \\
\indent $C_{>v}$ consists of constraints of the form  $(x > t + c')$, \\
\indent $C_{<c}$ consists of constraints of the form  $(x < c)$, and \\
\indent $C_{<v}$ consists of constraints of the form  $(x < t + c')$.\\
\\
Note $\backsim \in \{<, =\}$, and $c, c' \in \N$ are integer
constants, and $t \in T$ is a timing variable.

\medskip
If $|C_{=c}| > 1$, then $C_{=c}$ itself is unsatisfiable and so is
$C(A)$. Otherwise if $(x = c) \in C_{=c}$ then
check whether constraints in $C_{>c} \cup C_{<c}$ are
satisfiable on assigning $c$ to $x$. If not, then $C(A)$ is also
not satisfiable. Otherwise, $\forall t_1 \in T$ such that $(x = t_1
+ c_1) \in C_{=v}$, we can assign valuation $c - c_1$ for $t_1$; 
$\forall t_2 \in T$ such that $(x < t_2 + c_2) \in C_{<v}$, we can
assign valuation $(c - c_2) + z$, ($z >0: (c - c_2) + z  > 0$) for $t_2$; and $\forall t_3 \in T$
such that $(x > t_3 + c_3) \in C_{>v}$, we can assign $(c - c_3)
-z$, ($z > 0:(c - c_2) - z  > 0$) to $t_3$. Also assign $c$ to $y$ if $(x = y) \in C_{xy}$, else assign
$c+1$.

In the other case, when $C_{=c} = \emptyset$, calculate $l
= \max(C_{>c})$ if $C_{>c} \neq \emptyset$, else $l = -\infty$ and
$m = \min(C_{<c})$ if $C_{<c} \neq \emptyset$, else $m = \infty$. 
We define $ \max(C_{>c}) = \max \{ c \in \nR \,|\, x > c \in C_{>c} \}$, 
and $ \min(C_{<c}) = \min \{ c \in \nR \,|\, x < c \in C_{<c} \}$. 
Next we check if $l < m$. If not, these constraints cannot be
satisfied simultaneously. Otherwise we can choose any value of $x,\, l < x < m$,
as a solution. Satisfying valuations to all timing variables can
be assigned accordingly.

To estimate the time complexity, notice that partitioning of $C(A)$
can be done in linear time with respect to the size of the
constraint set since in order to place a constraint in its correct
partition it only requires to check the form of inequality and
type of variable (constant or variable). All other steps of
checking satisfiability and assigning valuations to timing
variables in $T$ can also be carried out in time linear on the
size of the constraint set, where size of the constraint set is
bounded by $|Cl(\phi)|$\hfill \bbox

\begin{Theorem}\label{satcomplexity}
Satisfiability problem for (unquantified) TLTL is PSPACE Complete.
\end{Theorem}
{\bf Proof.} Let $|\cA_\phi|$ denote the size of the structure $\cA_\phi$, which is bounded by the number of possible subsets of $Cl(\phi)$, that is, $|\cA_\phi| \leq 2^{\cO(|Cl(\phi|))}$. The number of constraints appearing in any atom are also bounded by $|Cl(\phi)| \leq 7|\phi|$, therefore $|\cA_\phi| \leq 2^{\cO(|\phi|)}$. By Lemma~\ref{lemmac}, consistency checking of these constraints can be performed in time $\cO(|Cl(\phi)|)$. This results in an overall time-bound $2^{\cO(|\phi|)}|\phi| = 2^{\cO(|\phi|+log|\phi|)} = 2^{\cO(|\phi|)}$.

Using a similar argument presented in~\cite{SC85}, we can conclude that there exists a nondeterministic algorithm $\mathit{M}$, which (generates $\cA_\phi$ `on-the-fly' and) accepts $\phi$ $\mathit{iff}$ it is satisfiable.

$\mathit{M}$ uses space of the order of $|Cl(\phi)|$. Using Savitch Theorem~\cite{Savitch70}, it can be concluded that there exists a polynomial space bounded $(\cO(|Cl(\phi)|^2))$ deterministic
algorithm which can decide satisfiability of a TLTL formulae.

It is also shown in~\cite{SC85} that satisfiability of LTL with $\unt$ and $\bigcirc$ is PSPACE-hard. Since LTL is properly embedded in TLTL, it renders satisfiability of (unquantified) TLTL PSPACE-complete. \hfill \bbox

As a consequence, we also have,

\begin{Theorem}\label{valcomplexity}
Validity problem for (quantified) TLTL is PSPACE Complete.
\end{Theorem}

\section{Model Checking for TLTL}\label{mc}

The model checking problem of deciding whether a TLTL formula $\psi$ is satisfied by all the computations of a given timeout program $P$ with clock, timeout, and static timing variables, is conceptually much harder than deciding the validity of TLTL formulas. This difficultly arises due to the fact that clock, timeout, and static timing variables range over the set of non negative reals, and therefore timeout systems are inherently infinite state systems. This render automated verification of these systems difficult as most of the model checking techniques proceed by exhaustive enumeration of the state space. 

Therefore we consider a restriction of TLTL over $\N$ (i.e., clock, timeout, and static timing variables assume positive integer valuations). Also we restrict our attention to only those timeout systems where increments in the values of the timeout variables and thus, the clock increments are allowed only over a finite range of values, while taking transitions. 

\subsection{Timeout Programs}\label{timeout-program}

The representation of a finite state timeout program that we consider, is given by a {\it timeout Kripke structure} (TKS) $K = \lan S, S^0, E \ran$ over the clock $x$, the set of static timing variables $T$, a finite set $\cT$ of timeout variables ${\tau}_1, {\tau}_2, \ldots, {\tau}_n$ used to record the values of timeouts such that $\cT \cap T = \emptyset$, and a variable $y$ which equals $\min\cT = \min\{{\tau}_i \,:\, {\tau}_i \in \cT\}$, where

\begin{itemize}
\item $S$ is a finite set of locations. Each location $s \in S$ gives a boolean interpretation to each of the propositions and an integer interpretation to static timing variables appearing in $\psi$ (i.e., the set $T_\psi$) in the interval $[0, M]$,
\item $S^0 \subseteq S$ is the set of initial locations defining the values for static timing variables for the runs starting from these locations,
\item $E = ( E^{+} \cup E^{0} ) \subseteq (S \times \N \times (\N \cup \{\star\}) \times S)$ - denotes the set of edges connecting locations in $S$. $E$ is partitioned into two disjoint sets $E^{+}$ and $E^{0}$. If $(s, l, m, s') \in E^+$ then $l = m = 0$. For simplicity we omit $l$ and $m$ for $E^+$ edges and represent them as $(s, s')$. For $E^0$
edges either $l$ and $m$ assume non zero positive integral values, which define the finite range of values for incrementing timeouts or, specifies open ended range of values larger than $l$ for incrementing timeouts when $m$ is $\star$.
\end{itemize}
The operational meaning to $E^+$ and $E^0$ is given as follows.
\begin{itemize}
\item $E^+$ is the set of delay transitions, whereby clock $x$ advances to $\min \cT$, that is, if $(s, 0, 0, s') \in E^+$, then on taking this transition, the value of the clock $x$ is incremented to $\min \cT$.
\item $E^0$ represents the set of the discrete transitions. For $(s, l, m, s') \in E^+$ on a discrete transition at least one of the timeouts attaining the minimum value is incremented by some arbitrary value $\delta$ in $[l, m]$ (if $m \in \N$), or $\delta \geq l$ (if $m$ is $\star$). 
\end{itemize}

The semantics of a TKS  $K$ is defined as follows: We define a timeout computation of $K$ to be an infinite sequence of timeout states $$\sigma : \lan s_0, x_0, y_0, \cT_0 \ran, \lan s_1, x_1, y_1, \cT_1 \ran, \cdots, $$ where $x_0, x_1, \cdots, $ denote the clock values, $y_0, y_1, \cdots, $ denote the values for the variable $y$, and $\cT_0, \cT_1, \cdots$ denote sets of values for the timeouts in $\cT$ for $i = 0, 1, \ldots$ such that $\cT_i[j]$ would denote the value of $\tau_j$ in $\cT_i$. All (static) timing variables in $T$ assume the same valuation in every state. Thus we have,

\begin{itemize}
\item $s_0 \in S_0$ and either $x_0 = y_0 = \min \cT_0 = 0$ or $0 = x_0 < y_0 = \min \cT_0$.
\item For every $i = 0,1, \ldots$ 
\begin{description}
  \item[-] $\forall t_j \in T: \, s_i(t_j) = s_0(t_j)$
  \item[-] $y_i = \min \cT_i$. 
\end{description}  

\item For every $i = 0,1, \ldots$ 
\begin{description}
  \item[-] either $(s_i, 0, 0, s_{i+1}) \in E^+$, s.t. $x_{i} < \min \cT_i \wedge x_{i+1} = \min \cT_i$. Also $\cT_{i+1} = \cT_i$, that is, during delay transitions timeouts do not change.
  \item[-] or $(s_i, l, m, s_{i+1}) \in E^0$ and $\exists \tau_j \in \cT$ s.t. $\cT_i[j] = \min \cT_i$, and $\cT_{i+1}[j] = \cT_i[j] + \delta$ where $\delta \in [l, m]$ if $m \in \N$, otherwise $\delta \geq l$ if $m$ is $\star$. Also $x_{i+1} = x_i = \min \cT_i$ and $\forall \tau_k \in \cT\setminus\{\tau_j\}. \cT_{i+1}[k] = \cT_i[k]$.
\end{description}
\item There are infinitely many $i's$ such that $x_{i+1} = \min\cT_i$,
which means clock and timeouts always advance.
\end{itemize}

\subsection{A Tableau Construction for the Product of the program $K$ and the formula $\phi$}\label{cross-product}

We construct a tableau $\cK = \cA_{\phi \times K}$ as the cross product of the tableau for a (unquantified) satisfiable TLTL formula $\phi$ and a TKS $K$. The elements of $\cK$ are
\begin{itemize}
\item $N_{\cK}$ is the set of the nodes consisting of pairs $\lan A,s \ran$ with $A \in {\cA}_{\phi}$ (tableau for $\phi$) and $s \in K$.
\item $E_{\cK} = E_{\cK}^+ \cup E_{\cK}^0$ is the transition relation where $E_{\cK}^+$ captures the elapse of time and
$E_{\cK}^0$ represents the discrete transition. Let $u::= t + c\;|\; c$, which is defined in Section~\ref{tltl-syntax}.
\begin{description}
	\item[-] $(\lan A,s \ran, \lan A', s' \ran) \in E_{\cK}^+$ iff $(A,A') \in R, (s,0,0,s') \in E^+$ and
		\begin{description}
			\item[] $x < u \in C(A) \Rightarrow x = u \in C(A')$ or $x>u \in C(A'),$ 
			\item[] $x = u \in \ C(A) \Rightarrow x > u \in C(A'),$ and 
			\item[] $x<y \in C(A) \Rightarrow x=y \in C(A')$
		\end{description}
	\item[-] $(\lan A,s \ran, \lan A',s' \ran) \in E_{\cK}^0$ iff $(A,A') \in R$ and $(s,l, m, s') \in E^0$ and
		\begin{description}
			\item[] $x \backsim u \in C(A) \Leftrightarrow x \backsim u \in C(A')$
			\item[] $x = y \in C(A) \Rightarrow$ 
			$(x < y) \in C(A')$~\footnote{All the timeouts with minimum value are incremented on taking the transition $(s,s')$.}
		\end{description}
\end{description}
\item $N_0$ is the set of initial nodes consisting of all pairs $\lan A, s \ran$ such that $\phi \in A$ and $s \in S_0$.
\end{itemize}

\subsection{Model Checking Procedure}

We check if all runs of a program $K$ satisfy a TLTL-formula $\psi = \forall t_1\ldots t_k. \phi$ as follows: 
\begin{description}
    \item[Step1] Construct the initial tableau ${\cA}_{\neg\phi}$ for the negated formula $\neg\phi$ as described in Section~\ref{validity}.
    \item[Step2] Construct the tableau product $\cA_{\neg\phi \times K}$ as described in the Section~\ref{cross-product}.
    \item[Step3] Check if ${\cA}_{\neg\phi \times K}$ contains a self-fulfilling path for $\neg\phi$.
\end{description}

\begin{Lemma} The TKS $K$ satisfies $\neg\phi$ if and only if ${\cA}_{\neg\phi \times K}$ contains a self-fulfilling path.\end{Lemma}
\begin{Theorem} The TKS $K$ validates the TLTL specification $\psi$ if and only if it does not satisfy $\neg\phi$. \end{Theorem}

\subsection{Complexity of Model Checking}

The size of the product tableau $\cA_{\neg\phi \times K}$ is bounded by $\cO(|K| \times |\cA_{\neg\phi}|)$ or $\cO(|K|\times 2^{7|\phi|})$, which is linear in the size of the TKS and exponential in the size of the TLTL specification $\phi$. Since deciding the presence of a self fulfilling path can always be done in the worst case in time linear on the size of the product graph, we conclude that \textit{the problem if a TLTL-formula $\psi=\forall t_1\ldots t_k. \phi$ holds in a TKS $K$ can be decided in deterministic time linear in the size of the $K$ and exponential in the length of $\phi$}. 

Following the argument presented for satisfiability checking in Theorem~\ref{satcomplexity}, there exists a non deterministic algorithm which checks if ${\cA}_{\neg\phi \times K}$ contains a self-fulfilling path for $\neg\phi$ using $O(|\phi|)$ space. This renders the model checking also in PSPACE. To check the hardness part, we need to reduce the validity problem for TLTL to model checking, which requires  defining a TKS $K$ of constant size such that formula $\phi$ holds $\mathit{iff}$ it is valid in $K$. Towards that, we further assume that the range of static timing variables are restricted to the interval $[0, M] \subseteq \N$, where the value of $M$ can be approximated by the maximum path delay in the Timeout Kripke structure defined below. The \emph{Path delay} for a specific (acyclic) path starting from some initial location and ending at some designated location is the sum of the maximal possible timeout increments or clock delays (replacing open ended timeout increments with arbitrary values) over the transitions across the path. Well-known shortest path algorithms~\cite[580–-642]{CLRS01}, {\it viz.}, Floyd-Warshall algorithm, Dijkstra's algorithm, can be easily be adapted for calculating such maximal path delay over a given TKS. Now, choose $K = \lan 2^{\cP \cup [0, M]},2^{\cP \cup [0, M]}, 2^{\cP \cup [0, M]} \times \{0\} \times \{\star\} \times 2^{\cP \cup [0, M]} \ran$ to be the complete graph over all subsets of $\cP \cup [0, M]$.

\section{Undecidability of Dense TLTL}\label{undecide}

We relax the time-progress condition and consider an interpretation of TLTL formulas over a dense time domain. We prove the resulting logic to be highly undecidable by reducing a $\Sigma_1^1$-hard problem to its satisfiability problem. 

\subsection{2-counter Machines}
A {\it nondeterministic 2-counter machine $\cM$} consists of two counters $C_1$ and $C_2$ assuming non negative integer values, and a finite sequence of labeled instructions (e.g., labeled by numbers $1, 2, \ldots$) 
Each instruction may either increment or decrement one of the counters, or jump, conditionally upon one of the counters being zero. When the machine $\cM$ executes a non-jump instruction, it proceeds non-deterministically to one of two specified instructions. For example, using programming pseudo-code notation, $j^{th}$ instruction may be either of the following, where $i \in \{1, 2\}$:
\begin{eqnarray}
j\ &:&\ C_i := C_i + 1;\ goto\ l_1\ \mbox{or}\ l_2, \label{eq2}\\
j\ &:&\ C_i := C_i - 1;\ goto\ l_1\ \mbox{or}\ l_2, \label{eq3}\\
j\  &:&\ if\ C_i = 0\ goto\ l_1;\ else\ goto\ l_2,  \label{eq4}
\label{eq:i-th-inst}
\end{eqnarray}
where $l_1$ and $l_2$ are instruction labels. The configurations of such a $\cM$ having $n \geq 0$ instructions are represented by triples $\lan i, c, d \ran$, where $0 \leq i < n$ is the instruction label, and $c \geq 0, d \geq 0$ are the current values of the counters $C_1$, and the counter $C_2$ respectively. 
The relation between consecutive configurations can be defined in an obvious way. 
A computation of $\cM$ is a $\omega$ sequence of related configurations, beginning with the initial configuration, which is usually taken as $\lan 0,0,0 \ran$. 
Importantly, 2 counter machines are Turing complete~\cite{HUM01}. 
For more details on counter machines see~\cite[Chap. 8]{HUM01},~\cite[Chap. 7-8]{Jones97}. 

The computation of a counter machine is called {\it recurring} if it contains infinitely many configurations with the value of the instruction counters being $0$. 
It was shown in~\cite{AH89} that the problem  of deciding if a given nondeterministic $2$-counter machine has a recurring computation is $\Sigma_1^1$-hard.

\subsection{Dense TLTL} 

Let us relax the time-progress condition ($m_2$) and extend the expressive power of TLTL by providing a dense semantics to it, {\it i.e.}, we assume that between any two given time points there is another time point. 
We assume our time domain as non-negative rationals $\nQ$ with dense linear order induced by usual `$<$' relation, which is irreflexive, comparability-permissible and transitive. 

The technique we use to prove the undecidability of dense TLTL follows closely the one described in~\cite[Section 4.4]{AH89} to prove similar result for TPTL. 
We need a successor function $\cS$ on the underlying time domain $\nQ$. This function, when applied to an element in $\nQ$ will return an unique element greater than the original element. 
$\cS$ satisfies the following axioms: i) $q < \cS(q)$ for all $q \in \nQ$ and, ii) $q < q' \Rightarrow \cS(q) < \cS(q')$ for all $q, q' \in \nQ$. 
Note that owing to the denseness of $\nQ$, arbitrary many time points could be squeezed into a finite interval with the application of successor. For notational convenience, $\cS(q)$ will be represented as $q^+$ in the following discussion. 

We encode a computation of $\cM$ by using propositions $p_0, p_1, \ldots, p_n, r_1$ and $r_2$, precisely one of which is true in any state. 
The configuration $\langle i, c, d \rangle$ of $\cM$ is represented by the finite sequence $p_i, \overbrace{r_1, \ldots, r_1}^{c}, \overbrace{r_2,\ldots, r_2}^{d}$ of states.

The initial configuration $\langle 0, 0, 0 \rangle$ can be encoded using a proposition $p_0$. 
The recurrence condition can be encoded as $(\Box \Diamond p_0)$. 
It is possible to have the $k$-th configuration of a computation of $\cM$ correspond to the finite sequence of states that is mapped to the interval $[t, t^+)$.
We force the time to increase by a strictly positive amount between each successive states using $\Box (x = t \Rightarrow \bigcirc(x > t))$. Now we can copy groups of $r$-states by establishing a one-to-one correspondence of $r_j (j = 1,2)$-states at time $t$ and time $t^+$. 
In the following we assume that $t_0,t_1,t_2,\ldots,$ stand for static timing variables. 

Let us consider the instruction (\ref{eq2}) $j:\ C_2 := C_2 + 1;\ goto\ l_1\ \mbox{or}\ l_2,$ which increments the counter $C_2$ and proceeds nondeterministically to either instruction $l_1$ or $l_2$. 
We can encode this computation by the following TLTL-formula: 
\[ \Box(\phi \Rightarrow (\psi_1 \wedge \psi_2 \wedge \psi_3(r_1) \wedge \psi_3(r_2) \wedge \psi_4^{r_2})), \] where
\[
\begin{array}{ll}
\phi: & x = t \wedge p_j\\
\psi_1: & \dia(x = t^+ \wedge (p_{l_1} \vee p_{l_2})) \\
\psi_2: & \Box( x = t_1 \wedge \bigcirc(x = t_2 \wedge x < t^+) \Rightarrow
\dia(x = t_1^+ \wedge \bigcirc(x = t_2^+)))\\
\psi_3(r_j): & \Box((x = t_3 \wedge x < t^+ \wedge r_j) \Rightarrow \dia(x = t_3^+ \wedge r_j))\\
\psi_4^{r_2}: & \Box((x = t_4 \wedge \bigcirc(x = t^+)) \Rightarrow \dia(x = t_4^+ \wedge \bigcirc r_2 \wedge \bigcirc\bigcirc (x = t^{++})))\\
\end{array}
\]
The formula $\phi$ specifies that the current state at time $t$ corresponds to instruction $j$. 
The first conjunct $\psi_1$ ensures the proper progression to one of the two specified instructions, $l_1$ or $l_2$ at time $t^+$. 
The second conjunct $\psi_2$ establishes a correspondence between states in successive intervals $[t, t^+)$ and $[t^+, t^{++})$ representing configurations while the formula $\psi_3(r_j)$ copies $r_j$-states in the corresponding states from first interval to the next. 
The last conjunct $\psi_4^{r_2}$ adds a $r_2$-state at the end of next configuration, as required by the increment operation. In case of counter $C_1$ getting incremented, we will have $\psi_4^{r_1}$ instead of $\psi_4^{r_2}$ specifying an addition of a $r_1$ state at the beginning of the $r$-state sequence in the next configuration: $$\psi_4^{r_1}: \Box((x = t \wedge \bigcirc(x = t_4 \wedge ((r_1\vee r_2)\vee(p_{l_1}\vee p_{l_2})))) \Rightarrow \dia(x > t^+ \wedge x < t_4^+ \wedge r_1 \wedge \bigcirc(x = t_4^{+})))$$ 

Next, for the instruction (\ref{eq3}) $j\ :\ C_2 := C_2 - 1;\ goto\ l_1\ \mbox{or}\ l_2$, which specifies a decrement operation on $C_2$, we copy all $r_1$ states as specified by $\psi_3(r_1)$ above. However we copy the $r_2$ states excluding the last copy in the sequence. This is achieved by first modifying $\psi_3$ for $r_2$ as follows: $$\psi_3'(r_2): \Box((x = t_3 \wedge x < t^+ \wedge r_2 \wedge \neg\bigcirc(x = t^+)) \Rightarrow \dia(x = t_3^+ \wedge r_2))$$ and then rewriting $\psi_4^{r_2}$ as $$\psi_{4'}^{r_2}: \Box((x = t_4 \wedge x < t^+ \wedge r_2 \wedge \bigcirc(x = t^+)) \Rightarrow (x = t_4^+ \wedge \bigcirc(x = t^{++})))$$In case of decrement on $C_1$, we copy all the $r_2$ states as specified by $\psi_3(r_2)$, however copy the $r_1$ states only after excluding the first copy in the sequence. This is achieved by modifying $\psi_3$ for $r_1$ as follows: 
\[
\begin{array}{ll}
\psi_3'(r_1): & \Box(\psi^{yes}_3 \wedge \psi^{no}_3), \mbox{where}\\
\psi^{yes}_3: & (x = t_3 \wedge x < t^+ \wedge r_1) \Rightarrow \dia(x = t_3^+ \wedge r_1)\\
\psi^{no}_3:  & \neg(x = t \wedge \bigcirc(x = t_3 \wedge r_1) \Rightarrow \dia(x = t_3^+ \wedge r_1))
\end{array}
\]
Finally, we encode the \textit{if-else} instruction (\ref{eq4}) $j\ :\ if\ C_1 = 0\ goto\ l_1;\ else\ goto\ l_2;$ as following: \[ \Box(\phi \Rightarrow (\psi_1' \wedge \psi_2' \wedge \psi_3'(r_1) \wedge \psi_3'(r_2)))\] where
\[
\begin{array}{ll}
\psi_1': & ((x = t \wedge \bigcirc(r_2 \vee (p_{l_1} \vee p_{l_2}))) \Rightarrow \dia(x = t^+ \wedge p_{l_1})) \\ 
 & \bigvee ((x = t \wedge \bigcirc(r_1)) \Rightarrow \dia(x = t^+ \wedge p_{l_2}))\\
\psi_2': & \Box( x = t_1 \wedge \bigcirc( (x = t_2 \wedge x < t^+) \Rightarrow \dia(x = t_1^+ \wedge \bigcirc(x = t_2^+))))\\
\psi_3'(r_j): & \Box((x = t_3 \wedge x < t^+ \wedge r_j) \Rightarrow \dia(x = t_3^+ \wedge r_j))\\
\end{array}
\]
In case of $j\ :\ if\ C_2 = 0\ goto\ l_1;\ else\ goto\ l_2$, we modify $\psi_1'$ as follows: $$\psi_1'': ((x = t \wedge \neg\dia(r_2)) \Rightarrow \dia(x = t^+ \wedge p_{l_1})) \bigvee ((x = t \wedge \dia(r_2)) \Rightarrow \dia(x = t^+ \wedge p_{l_2}))$$
Thus for this 2-counter machine, $\cM$ we can construct a formula $\phi_{\cM}$ such that $\phi_{\cM}$ is satisfiable iff $\cM$ has a recurring computation. 
Hence the satisfiability of TLTL is $\Sigma_1^1$-hard.

We observe that the satisfiability of a TLTL formula $\psi$ can be always expressed as a $\Sigma_1^1$-sentence implying the existence of a model for $\psi$. 
Since $\nQ$ is countable, $\psi$ will also have a countable model. Thus any state sequence $\sigma$ for $\psi$ can be encoded by finitely many infinite sets of natural numbers in first-order arithmetic; say, one for each proposition $p$ in $\psi$, characterizing the states in which $p$ holds. It is easy to see $\psi$, as a first-order predicate holds in $\sigma$. 
We conclude that the satisfiability of TLTL formulas is in $\Sigma_1^1$.

\begin{Theorem}
The satisfiability problem for dense TLTL formulas is $\Sigma_1^1$-complete.
\end{Theorem}

\section{Discussion}\label{discussion}

While existing real-time logics e.g., TPTL~\cite{AH89} can specify clock based dense time properties, TLTL is more suitable for expressing properties of timeout based real-time models for the given semantic interpretation using timeout dynamics where granularity of time is defined in terms of timeout updates. As discussed in Section~\ref{mc}, the infinite state space models of real-time
systems can be model checked over discrete time TLTL using the proposed abstractions on the Kripke structure.

Though we only consider minimum of the timeout values using a dummy variable $y$, dynamic constraints involving individual timeouts (e.g., constraints of the form $x \leq \tau_j + c$, where $\tau_j
\in \cT, c \in \N$) can be easily included in the vocabulary of the logic because the existing tableaux procedure presented in Section~\ref{tab-construct} can be seamlessly extended using the fact that in any state $s$ it is the case, $\forall \tau_j \in \cT. s(\tau_j) \geq s(y)$. Similarly extending the logic with constraints involving congruences similar to TPTL and arithmetic expressions involving more than one timing variables similar to XCTL would enhance the expressive power of the logic. Digitizability~\cite{HMP92} is yet another important property for applying discrete time verification techniques on dense time logics and models.  Quite often, not all the formulas in dense time logics are digitizable, thus not amenable to discrete time verification. It remains to be seen which fragment of TLTL is digitizable. We conclude by trying to compare  
TLTL with Monadic Second Order Logic of Order (MSO). It will be a routine exercise to show that TLTL 
can be embedded in MSO, following the work~\cite{AH93}. However as a future work, it would be interesting to characterize the fragment of MSO, for which TLTL will be expressively complete.

\bigskip
\noindent
{\bf Acknowledgment} Both the authors did this work when they were with HTS Research, Bangalore, India.

\bibliographystyle{alpha}
\bibliography{tltl}

\end{document}